\documentclass[twocolumn,showpacs,preprintnumbers,amsmath,amssymb]{revtex4}

\usepackage{dcolumn}
\usepackage{graphicx}

\begin{document}

\title{A simple minded question: \\
       Do we live in the four-dimensional spacetime?}

\author {P.\ Doleschall}
\affiliation{Wigner Research Centre for Physics, HAS, \\
             H--1525 Budapest P.O.B. 49, Hungary \\
             E-mail: doleschall.pal@wigner.mta.hu}

\begin{abstract}
A new approach in the Newtonian space and time, based upon the assumption
that the inertial mass is the quantitative measure of the matter, is
considered. It has been shown that in case of a special physical system,
a supposed matter transfer may reproduce the relativistic mass increase
of an accelerated particle. As a consequence, the relativistic time
dilation in the accelerated system can be explained. It is shown that
subsequent accelerations produce systems, whose equivalent masses and the
rate of the equivalent clocks do not depend on their relative velocity in
a way predicted by the theory of special relativity. It is also shown that
adding a small velocity dependent component to the static gravitational
force, the measured peri-helium shift and the light deflection can be
reproduced. Some explanations of the experienced behaviour of the photons,
based on their changing mass, are offered.
\end{abstract}

\pacs{0.155.+b}

\maketitle

\section{Preface}

The author of the present paper is not sure that the following
ideas are worth for publication or they are more similar to a
somewhat lunatic conception. Besides that the author has to apologize
for the incompleteness of the presented old-fashioned phenomenological
approach, especially that it wishes to arise some doubt about the
absolute necessity of the present-day universally accepted
four-dimensional spacetime. This is a delicate and rather unpleasant
situation, since in the last hundred year everything, what belong to
the mainstream of the modern physics, was modelled in the
four-dimensional spacetime. Therefore it is clear, that based on an
automatic reflex, the presented approach will be rejected. On the other
hand, it has to be admitted that there may be somewhere a trivial error
in the presented approach, some experimental evidences may prove that
it cannot be right or because its embryonic state it may come to a
deadlock.

Therefore the author asks an emphatic patience from the readers willing
to consider the presented ideas.


No detailed references are given, because the present approach needs only
the basic knowledge of the Newton's \cite{NEW} and Einstein's \cite{EIN}
physics. All other informations are available on text-book level.

\section{Introduction}

Although most physicists hate the philosophical approach, a basic
hypothesis, the concept of the matter, is used. In natural sciences
the matter (whatever it is) forms our universe. The matter is supposed
to exist exclusively, it cannot diminish and cannot come into being
from nothing.

This concept of the matter can be used only in case, when one can find
a measurable physical entity characterizing the quantity of the matter.
This physical entity expected to be a scalar one and the same for any
observer. If isolated material systems (systems without a material
connection) exist at all, then the quantity of their matter content,
and consequently the physical entity characterizing it, must remain
constant.

According to the Newtonian concept, the matter supposed to exist in
the three-dimensional space and the change of its distribution (or the
change of its position if one perceives it as a body) is characterized
by the time. The Newtonian space and time are independent on each other.
The measures of the distances in the space and the length of the time
of course also has to be defined. The distance in a given system is
defined by using the extension of a piece of matter, which is believed
to have a constant size. The time, similarly to the distance, is
defined by a motion of a piece of matter, which is also believed to
be a stable constant in the given space. In this way all of the units
of the above introduced physical entities (quantity of matter, distance,
and time) are defined by their measurements.

This Newtonian frame (space and time) of the observations was challenged
by Einstein, stating that the observed world has to be described in the
four-dimensional spacetime, in which the measured space and time
are not independent on each other anymore. At present the spacetime
is considered the only correct frame for the description of physical
processes.

The instinctive perception that the world is composed of distinct
bodies was accepted by both of the Newton's and Einstein's model.
While in Newton's classical mechanics the distinct bodies may
interact with each other, generating their motion, the Einstein's
theory of special relativity based on the motion of the bodies in
the four-dimensional spacetime, without treating the reason of their
motion. The basic formulation of both models describes the motion
of pointlike material bodies (particles).

Although for the first glance it seems to be a nonsense, in the present
paper an attempt is made to return to the Newtonian pre-relativistic
physics. However, to attempt to do this, one must be able to describe
at least the most important basic phenomena which are treated as the
proofs of Einstein's theory and which were believed to be unexplainable
within the framework of the original Newtonian physics.

The first basic problem of the original formulation of the Newtonian
physics may be the insufficient concept of the force. It was defined
as an instantaneous and local one, whose origin is the interacting
bodies themselves. In that sense the force has a material origin,
but the force itself was not handled as a material one. However, if
our world is exclusively material, then the interaction between the
bodies also must be material and because of that, it must have a
matter content. This means, that a system of interacting particles
must be a distribution of some kind of matter in a certain space,
in spite of the fact that an observer perceives them as distinct,
although interacting particles. The extension of this matter
distribution in space may be a limited one: in this case the matter
of the system of interacting bodies occupies a finite volume of the
space. If the surrounding space is a vacuum (at present the classical
concept of vacuum is used: space without matter), then the system is
an absolutely isolated system, although one cannot be sure, that
such system may exist in the real world. A weaker definition of an
isolated system would be the one, which does not interact with its
surrounding. Such system can be called a semi-isolated system.
This is in most cases an approximation: the interactions within
the system are much stronger than the interaction between the
matter of the system and its surrounding. It seems to be evident
(although for a semi-isolated system the following statement is
an approximation) that the quantity of the matter of an isolated
system is conserved. If the Newton's first law is accepted, than
the total momentum of an isolated system must be conserved too.

The description of the physical processes of an isolated system,
which is in principle a matter distribution, must be done in the
space, which is occupied by its matter. Let us suppose that the
description of the system may be approximated as distinct bodies
and a common force-field, which is the source of the interactions
between the bodies. If the bodies are moving, one has to suppose
that their common force field is also changing. Therefore the
interaction between the bodies is expected to be a nonlocal
and time-dependent one, and what is worse, one cannot be sure
that the matter content of the perceived bodies remain the same
throughout their interacting processes. The additional problem is
that for the description of such a system one ought to find a
proper system of equations, which reproduces the observed behaviour
of such a material system. It seems to be evident that the proper
description of a matter distribution may be done only with some
kind of field theory.

\section{An alternative approach}

Even the simplest system, perceived by an observer as an isolated
system of two interacting bodies, is practically impossible to
describe properly on the basis that it is a matter distribution.
Therefore the obvious question arises: how is it possible that
the Newtonian physics is so successful, even in cases of systems
composed of a large number of interacting bodies. To answer this
absolutely justified question, let us scrutinize the Newton's
second law on the basis of the exclusively material world.

The Newton's second law has the following present-day form:
\begin{equation}
\frac{{\rm d} (m\vec v)}{{\rm d} t} = \vec F  \,\,\, ,  \label{e1.1}
\end{equation}
where the velocity $\vec v$ is the time derivative of the position
$\vec r$ of the pointlike body in a three-dimensional coordinate
frame: $\vec v={\rm d} \vec r/{\rm d} t$. The $\vec F$ is the force
generating the motion of this body with a mass $m$, which is the
measure of the inertia of the body. The force in Newton's theory
is supposed to be a local one and it acts instantaneously.

\subsection{A solvable two-body system}

The Newton's equation (\ref{e1.1}) does not include the body
(or bodies), which is supposed to be the source of the force.
In fact the Newton's equation describe a pointlike body moved by
a local and instantaneous force. This corresponds the following
isolated two-body system: i) one of the bodies has a large mass
$M$, which can be regarded as an infinite one relative to the
mass $m$ of the other lighter body; ii) the lighter body can be
treated as pointlike particle interacting with the heavy mass.
If the masses are handled as the quantitative measure of the
matter (this is the tacit view of the Newtonian physics) and in
the initial state the bodies have masses $M_{in}$ and $m_{in}$,
their sum $M_{in}+m_{in}$, because of the matter conservation law,
must remain constant throughout their interaction. The heavier,
practically infinite, mass has a practically fixed position in
the space and the lighter particle is moved by the force, generated
by the standing, infinite mass body. Of course the matter content
of the force-field, the source of the interaction, is part of the
full mass $M_{in}+m_{in}$ of the system. However, there is a
necessary extra condition, which enables one to use the Newton's
equation (\ref{e1.1}) for the description of such a system: one
has to suppose that the force-field, generated by the infinite
mass body, also has an infinite matter content. In this way, as a
first approach, one may suppose that the infinite mass force-field
of the infinite mass body remains practically unchanged during the
interaction process and therefore the force can be treated as a
local and instantaneous interaction acting on the lighter body.
In this way the Newton's equation (\ref{e1.1}) may be applied,
in spite of the fact that a general two-body system is expected
to be described as a system of matter distribution.

It has to be emphasized that the expression ''infinite mass'', used
for the mass of the heavier body, is just handled as an approximation:
the heavier mass is treated as a relatively infinite mass compared to
the mass of the lighter body (for example an elementary particle in
an accelerator). The coordinate frame of the description is fixed to
the center of mass of the isolated two-body system (in case of an
accelerator-particle system it is evidently fixed to the practically
standing accelerator). This ''infinite mass'' system of the above
described two bodies of course may move with a constant velocity
relative to other isolated systems.

Now let us apply the Newton's equation for the above described
special isolated system. First rearrange the equation (\ref{e1.1})
as ${\rm d}(m\vec v)= \vec F {\rm d} t$. Multiplying it with
$\vec v$, one gets the following equation:
\begin{equation}
\vec v {\rm d}(m\vec v) = \vec F {\rm d}\vec r \,\,\, ,  \label{e1.2}
\end{equation}
since $\vec v {\rm d} t={\rm d} \vec r$. In the present paper the
equation (\ref{e1.2}) is called the modified Newton's equation. 

One has to recognize that the quantity $\vec F {\rm d}\vec r$
corresponds to a small quantity of work performed by the force
$\vec F$ and as a such one, it is a small quantity of energy. If
the mass $m$ remains constant (this is supposed by the Newton's
theory), then the equation (\ref{e1.2}) becomes ${\rm d}(mv^2/2)=
\vec F{\rm d}\vec r= {\rm d}E_K$, which expresses that the quantity
of the work $\vec F {\rm d} \vec r$, due to the action of the force
$\vec F$ upon the particle, is equal to the change of the kinetic
energy $E_K$ of the particle.  

But what happens, if one does not suppose the mass $m$ to be a
constant, as it is assumed in Newton's theory?

\subsection{The mass}

In Newtonian physics the motion of the body is described in the
absolute space as a function of the absolute time. As it is
mentioned before, the inertial mass $m$ of the body is the tacit
measure of the quantity of the matter content of the body. This
Newtonian measure of the matter content of a body is scalar and
independent on the coordinate frame observed from, as it is expected
on the basis of the hypothesis of the exclusively material world.
This Newtonian inertial mass as a measure of the matter content of
a body has been so successful that even today this measure is used
in all other natural sciences, in spite of the fact that the theory
of relativity (if it is not stated differently, the present paper
refers to the theory of special relativity), discarded it as a
generally applicable measure for the quantity of matter.

Now let us forget for a moment the Einstein's theory and make the
crucial assumption: the Newtonian inertial mass is the physical
entity characterizing the quantity of all kind of matter.

This seems to be plausible, because when a particle is accelerated
by a force-field, its inertial mass includes both of the matter
contents of the particle and its force-field (if one may separate
them at all) and one accelerates both of them. If a particle is
accelerated to a constant velocity and it becomes a separate
isolated system, it will sustain its force-fields (for example the
gravitational or Coulomb field).

The assumption, that the Newtonian inertial mass is the quantitative
measure of the matter content of a body (which includes of course
the matter content of the force-field connected with the body), is
the basis of the present alternative approach.

Let us examines the consequences of this assumption.

If the mass is the quantitative measure of the matter content of
a particle, then the experienced phenomenon of the increased mass
of the accelerated particle by a relatively infinite mass force-field
must be an increase of its matter content. Since the present approach
is based upon the idea that a finite mass particle (with its own
force-field) interacts instantaneously with an infinite mass local
force-field, it seems to be evident that a matter (mass) transfer
must go on between the matter content of the particle and the matter
content of the infinite mass force-field.

The right-hand side of the equation (\ref{e1.2}) expresses a small
quantity of transferred energy $\vec F {\rm d}\vec r$, which ought
to have a matter (mass) content. The question is, how this matter
content, transferred from (or to) the infinite mass force-field can
be found. One way to do it to accept Einstein's famous result that
the full energy $E$ of a particle is expressed via its mass as
$E=c^2m$ ($c$ is the universal speed of the light in vacuum). But
this relation can be used also as an experimentally proved feature:
in a given system the full energy $E$ of an accelerated particle is
$c^2 m$ or $\sum_i c^2 m_i$ for a composite system. Therefore one
may suppose that the small amount of transferred energy (work) may
be expressed as $c^2{\rm d}m$, where ${\rm d} m$ is the transferred
mass. Consequently the following relation may be valid:
\begin{equation}
\vec F {\rm d} \vec r=c^2 {\rm d} m \,\,\, . \label{e1.3}
\end{equation}

The relation (\ref{e1.3}) is based upon the assumption that the
mass is not only a measure of the inertia, but it is a general
quantitative measure of the matter content both of the particles
and the force-fields. This is an extension of the Newtonian physics,
because the Newton's assumption, that the interacting bodies have
constant masses and consequently a constant matter content, is
discarded. In the present approach an energy transfer supposed to
be also a matter (and consequently a mass) transfer. The original
masses $M_{in}$ and $m_{in}$ are changing to values $M$ and $m$ with
the condition $M+m=M_{in}+m_{in}$. Since both the $M$ and $M_{in}$
are practically infinite compared to the mass of the lighter body,
the effect of a possible energy (and mass) transfer is seen only on
the lighter particle.

Using the equality (\ref{e1.3}), the modified Newton's equation
(\ref{e1.2}) can be written as
\begin{equation}
\vec v {\rm d}(m\vec v) = c^2 {\rm d} m \,\,\, .  \label{e1.4}
\end{equation}
Multiplying the equation by $m$ one gets the trivial solution
\begin{equation}
m = m_{in} \sqrt \frac{1-v_{in}^2/c^2}{1-v^2/c^2} \,\,\, ,
    \label{e1.5}
\end{equation}
where $m_{in}$ and $v_{in}$ are the mass and speed, respectively,
of the particle in its initial state. If the initial speed $v_{in}$
is the result of an acceleration of the particle at rest with a mass
$m_0$ by an infinite mass force-field of the same system or $v_{in}=0$,
then $m_{in}=m_0/\sqrt{1-v_{in}^2/c^2}$ and therefore
\begin{equation}
m = m_0 /\sqrt{1-v^2/c^2} \,\,\, . \label{e1.6}
\end{equation}
This is exactly the mass increase of the accelerated particle,
predicted by the theory of relativity, proven by measurements.

One may conclude that a special system has been found for what
i) the modified Newton's equation (\ref{e1.2}) seems to be valid;
ii) the assumptions that the mass is the quantitative measure of
the matter and the full energy of a piece of matter $m$ is $c^2m$,
lead to the solution (\ref{e1.5}), what reproduces the relativistic
dependence (\ref{e1.6}) of the mass of the accelerated particle at
rest in its initial state.

This result also supports that if the energy of the particle in
its initial state at rest (which may be treated as the internal
energy of the particle at rest) is $E_0=c^2m_0$, then the increase
of the full energy $E$ of the particle can be expressed by
integrating the equation (\ref{e1.3}) and the full energy of the
accelerated particle in its final state becomes $E=c^2m$. This
shows, that at least for this physical system, the equations
(\ref{e1.2}) and (\ref{e1.4}) are consistent with Einstein's
expression of the full energy of an accelerated particle.

However, comparing the relativistic mass dependence of a particle
with the present velocity dependence ((\ref{e1.5}) or (\ref{e1.6})),
the essential difference is that the present change of the mass,
as a quantitative measure of the matter content, is a real physical
process. Therefore the change of the mass i) is independent on the
coordinate frame it is observed from; ii) and is valid at every
moment of the process of the acceleration.

Contrary to that, the theory of relativity handles the increased
mass of an accelerated particle as a phenomenon valid only in the
accelerator's inertial frame. If the acceleration of the particle
is finished and it will reach a constant velocity $\vec V$, it
becomes an isolated system and according to the theory of relativity
for an observer in the inertial coordinate frame, fixed to the
particle, the mass of the particle will be equal to the rest mass
of the particle before the acceleration.

The concept of the inertial frame in both of the Newton's and Einstein's
theories is independent on the matter, it is just related to kinematics.
The basic difference is, that the equivalence of all inertial frames in
Newton's physics is a consequence of the Newton's laws (including the
assumption that the bodies have constant masses), while in Einstein's
theory it is postulated.

In the present approach the concept of the inertial frames are not pure
kinematical ones, they are interpreted as coordinate frames fixed
to isolated material systems moving relative to each other with a
constant velocity. The problem is, that if the change of the mass of an
accelerated particle is a real physical process, then the principle of
the equivalence of all inertial frames is questionable.

Now one may try to answer the question: how is it possible, that the
Newtonian physics is so successful? The reason is rather simple: in
most processes, described successfully by the Newtonian physics, the
velocities of the bodies are much less than the velocity of the light
$c$. This is a consequence, according to the present approach, that
the classical physics generally handles relatively weakly interacting
distinct bodies, whose full mass is believed to be much larger than the
mass content of their force-fields generating their interactions.
Therefore the masses of the perceived bodies can be approximated as
constant ones and the interactions between them, due to their low
speed, may be approximated as local and instantaneous ones.

However, these conditions are not valid for the world of atoms,
nuclei, and elementary particles. This can be the reason of the
necessity of the quantum theories, although the quantum mechanical
descriptions are also based upon the Newtonian pointlike approximation
of the particles with constant masses, which does not seem to satisfy
the requirement of the material character of the force-fields.

To illustrate the above statement, let us estimate the mass of the
Coulomb force-field of a free proton or electron (they are supposed to
be nearly the same). The proton, according to our present knowledge,
has a so called Coulomb radius less than $1 fm$. In the vacuum the
energy of the Coulomb field outside of a charged sphere with a radius
$R$ is:
\begin{eqnarray*}
E_{Coul}=\frac{1}{2}\int_R^{\infty} \frac{e^2}{r^4} r^2{\rm d} r=
         \frac{1}{2}\frac{e^2}{R} \,\,\,  ,
\end{eqnarray*}
where $e$ is the charge of the electron. At $R=1 fm$ this energy is
approximately $0.72 MeV$. The full energy of an electron at rest
($E=c^2m_e$) is approximately $0.511 Mev$, which is less than the
above calculated energy of the Coulomb field ($0.72 MeV$). It
seems to be clear that the full energy of an electron at rest is
comparable with the energy of its Coulomb field.

It has to be emphasized again that the velocity dependence of the
mass of a finite mass body, based on the above described modified
Newtonian model, can be calculated only in case, when the accelerator
has a relatively infinite mass force-field compared to the mass of the
accelerated body. Throughout the present paper, unless it is stated
differently, acceleration of pointlike finite mass systems (or bodies) by a
relatively infinite mass force-field is described in an Euclidean
coordinate frame, fixed to the relatively infinite mass.

\subsection{The time}

When the basic physical concepts (quantity of the matter, the distances
in the space, and the time) were introduced, it was emphasized that one
has to find measurable quantities characterizing these entities. The
time $t$, used in the Newton's second law is defined as a pointlike
quantity on a linear scale, whose origin is arbitrarily defined (the
$t=0$ point). Another possibility is to use the difference of two
points of time $t_1$ and $t_2>t_1$, which is the length of the time
or the period: $\tau=t_2-t_1$. In the next considerations the period
form $\tau$ is used as the concept of time.

Since the Newton's idea of absolute space and time cannot be defined,
within the present approach, the ''absolute'' space and time is defined
in the chosen system of the relatively infinite mass. All derivations
is done with this time. This means that the events in the moving system
is supposed to be observed from the accelerator's frame.

\subsubsection{The time in an accelerated system}

Let us investigate a subsystem, accelerated as a pointlike one to
a constant speed $\vec V$. Let us further suppose that before the
acceleration there was a gyroscope at rest in the subsystem with
a moment of inertia $I$, rotating with an angular velocity $\omega$.
After the acceleration the gyroscope remains in a rest state relative
to the accelerated subsystem and its mass is increased by a factor
$1/\sqrt{1-V^2/c^2}$. Because the gyroscope was also accelerated as
a pointlike body, its angular momentum, according to the Newtonian
physics, must be conserved. However, the size and shape of the
subsystem, together with the gyroscope, in principle may be changed
during the acceleration.

Now let us make an important assumption: suppose that the size and
shape of the subsystem and its parts remains unchanged during its
acceleration as a pointlike body. In this case the moment of inertia
$I'$ of the accelerated gyroscope is expressed by the formula
$I'=I/\sqrt{1-V^2/c^2}$. Supposing that the Newtonian rule of the
momentum conservation is true (there was no force to act on the
rotation of the gyroscope, otherwise one cannot suppose the system
to be pointlike), then the equality $I\omega=I'\omega'$ ought to be
true. This leads to the relation of the angular velocities of the
original and accelerated gyroscope: $\omega'=\omega \sqrt{1-V^2/c^2}$.
Using the more common frequency $\nu$ ($2\pi \nu= \omega$):
\begin{eqnarray}
\nu'=\nu \sqrt{1-V^2/c^2} \,\,\,  .  \label{e3.1}
\end{eqnarray}

The rotational energy $E_I=I\omega^2/2$ is also an important quantity
in the Newtonian physics. This may be regarded as the internal energy
of the gyroscope. Contrary to the Newtonian physics, where this is a
conserved quantity because of the supposed constant masses, the
increased mass of the accelerated gyroscope leads to the change of
this rotational or internal energy: in the accelerated system it
becomes $E'_I=I'\omega'^2/2$. Using the relation (\ref{e3.1})
it can be expressed as
\begin{eqnarray}
E'_I=E_I \sqrt{1-V^2/c^2} \,\,\, . \label{e3.2}
\end{eqnarray}
The internal energies $E_I$ and $E'_I$ are also supposed to be
independent on the system observed from.

Although angular velocity of an ideal gyroscope can be used to measure
the time in a system, the mechanical clocks are using the oscillating
motion of a cylindrical body with a certain moment of inertia $I$ for
measuring the period of the time. The speed of the run of a given
clock is inversely proportional to the moment of inertia of the given
cylindrical part. Therefore a clock accelerated to a constant speed
$\vec V$ runs $\sqrt{1-V^2/c^2}$ more slowly, which means that if
one defines the period $\tau'$ as the period of the time shown by
the accelerated clock, it is $\sqrt{1-V^2/c^2}$ times less than
the the same period $\tau$, shown by the original clock. Therefore
the periods $\tau$ and $\tau'$ are related as
\begin{eqnarray}
\tau'=\tau\sqrt{1-V^2/c^2} \,\,\,  .  \label{e3.3}
\end{eqnarray}

The formula (\ref{e3.3}) expresses the same time dilation than
that of the theory of relativity. However, contrary to the theory
of relativity, within the present approach this period $\tau'$ is
as real as the increased mass $m'$ and it must be also independent
on the system observed from.

If one supposes that the internal processes of an accelerated composite
elementary particle become slower, like the rotation of the gyroscope,
measured with the time, defined in the accelerator's system, then
one may also suppose that its longer mean lifetime is a consequence of
these slower inside processes. To generalize it, one may say that if
a composite body is accelerated to a constant velocity $\vec V$, its
internal processes, expressed with the units of the accelerator's
time, become slower by a factor of $\sqrt{1-V^2/c^2}$.

In order to reproduce the longer mean lifetime of an accelerated
decaying particle, the present approach has to suppose that the
acceleration does not change the size and shape of the gyroscope.
Therefore this assumption has to be added as a new basic assumption
to the present approach, otherwise the simplest form of the time
dilation, the change of the mean lifetime of a decaying particle,
cannot be explained.

In a next step one may try to generalize the behaviour of the
internal energy of an accelerated gyroscope. Such generalizations
are rather common in physics. Let us suppose that the internal
energy $E'_I$, which is defined as the internal energy of a body
at rest in the pointlike system, accelerated by an infinite mass
force-field to a constant speed $\vec V$, behaves as the internal
energy of the gyroscope: $E'_I=E_I \sqrt{1-V^2/c^2}$, where $E_I$
is the internal energy of the body before the acceleration. This
internal energy may be an internal motion or it may be a capacity
to generate motion between the parts of the body. In that sense
it must be also independent on the system observed from, like
in case of a gyroscope.

The consequence of the generalization (\ref{e3.2}) is that the case
$V \rightarrow c$ leads to the $E'_I \rightarrow 0$. This means that
in this limit an accelerated clock practically stops, which is
equivalent to the practical stop of the internal motions in the
accelerated system. If one believes that the internal energy is a
capacity to generate motions within the accelerated body, in case
of $V \rightarrow c$ this capacity diminishes. The $V \rightarrow c$
produces a kind of a ''frozen'' state.

Now let us choose a subsystem at rest in an infinite mass system,
which includes an electron-positron pair in a state at rest in
the coordinate frame fixed to the subsystem. Although this two-body
system of the electron-positron pair does not fit to the systems,
which can be described within the framework of the present approach,
if one supposes that it is a semi-isolated system within the
subsystem in both its initial and accelerated states, then using
only their initial and final states, certain statements can be done.

The measurements show that if this electron-positron pair annihilates
in the original system, photons arises with masses $m_f$ (in most
cases only two photons) and because of the momentum and energy
conservation, the relations $\sum_f m_f \vec c=0$ and $c^2\sum_f m_f=
2c^2 m_e$ ($m_e$ is the identical mass of the electron and positron)
are fulfilled. In the present approach the energy conservation is
equivalent to the matter (mass) conservation: $\sum_f m_f=2m_e$. Since
the expression $E=c^2 m$ of the theory of relativity for the full
energy of a particle is also accepted, the electron and positron at
rest in their initial states have an internal energy $E_I$ equal to
their full energy: $E_I=c^2 m_e$. Therefore in the annihilation
process the full energy of the arising photons is equal to the sum
of the internal energies of the electron and positron.

Now the subsystem together with the electron-positron pair is
accelerated to a constant speed $\vec V$. If the generalization
expressed in relation (\ref{e3.2}) is true, the internal energy
$E'_I$ of the accelerated electron and positron becomes $E'_I=
E_I\sqrt{1-V^2/c^2}=c^2 m_e\sqrt{1-V^2/c^2}$. Let us suppose
that the electron-positron pair annihilates in the accelerated
subsystem in the same way as it happened in the original system
(the system of the accelerator), namely that all arising photons in
the coordinate frame fixed to the accelerated subsystem have the
same velocity $c'_f=c'$, independent on their directions, and their
energies are $c'\,^2 m'_f$. Here it has to be emphasized again, that
the velocity $c'$ is the speed of the photons in the accelerated
subsystem, seen from the original system. It is a derivation of the
distance in the subsystem (supposed to be the same as the corresponding
distance in the original system) by the time of the original system.

In the present approach the mass conservation gives a relation
$\sum_f m'_f=2m'_e$. The source of the full energy of photons
in the coordinate frame of the accelerated electron-positron pair
must be the internal energy of the accelerated electron-positron
pair: $c'\,^2 \sum_f m'_f=2E'_I=2c^2 m_e\sqrt{1-V^2/c^2}$.
Because the $\sum_f m'_f=2m'_e=2m_e/\sqrt{1-V^2/c^2}$, one
gets the solution
\begin{eqnarray}
c' = c \sqrt{1-V^2/c^2} \,\,\,  . \label{e3.4}
\end{eqnarray}

This means, that in the present approach the speed of the photon in the
accelerated system where it is generated, would have a velocity less
than the speed of the photon $c$ in the original system. The key concept,
which leads to a smaller speed of the photon in the accelerated system
is the supposed decrease of the internal energy of the accelerated
particles. This would not be enough without supposing that the annihilation
process of the accelerated electron-positron pair obeys the same rules
than those of the original system. The internal energy of an accelerated
electron at rest in the coordinate frame, fixed to the accelerated system,
is  $E'_I=c'\,^2m'_e$. This is the generalization of Einstein's
$E_I=c^2m_e$ formula and equivalent to the relation (\ref{e3.2}).

If one would like to compare the photons arising from the annihilation
of an electron-positron pair in the original and in the accelerated
system, one may try to use the rules given by quantum mechanics. For
the simplicity let us assume that only two photons are emitted. In
this case
\begin{eqnarray*}
m_e c^2 = h \nu \: \: \: , \: \: \:  m'_e c'\,^2 = h \nu'
   \,\,\, ,
\end{eqnarray*}
where $h$ is the Planck constant.
Using the expressions (\ref{e1.6}) and (\ref{e3.4}), the above
quantum mechanical expressions reproduce the expression (\ref{e3.1}):
$\nu'=\nu \sqrt{1-V^2/c^2}$. This means that the photons emitted in
the annihilation process of an electron-positron pair of a system
moving with a constant speed $\vec V$, have a frequency
$\sqrt{1-V^2/c^2}$ times less, compared to the photons emitted
in the original system. If the rate of the clock of the accelerated
system is regulated by this frequency, one reproduces the slowdown
of the accelerated clock ((\ref{e3.1}) and (\ref{e3.3})).

The next case is the frequency of the light, emitted by a certain atom
in the system at rest. The transition frequency of the emitted photon
is proportional to the energy difference between two given energy
levels of the given atom. It is easy to realize, that the energy levels
of an atom are the levels of its internal energy. If the atom is
accelerated to a constant speed $\vec V$, its mass increases and
consequently its internal energy decreasing according to the supposed
rule (\ref{e3.2}). The same change is valid for its energy levels.
Therefore in the accelerated system the transition frequency between
two given energy levels will be less by the factor $\sqrt{1-V^2/c^2}$
(see (\ref{e3.1})).

Therefore if the rate of the clock is based upon the frequency of given
photons (at least in the two elaborated cases), the clock accelerated
to a constant speed $\vec V$, will run $\sqrt{1-V^2/c^2}$ times slower
than the original standing clock. This is the same result what was
found in case of a mechanical clock. In fact the frequency of a given
photon is just the physical quantity, which is presently used to measure
(and in this way to define by the measurement) the time.

In order to reproduce the slower rate of the accelerated clock the
following additional assumptions had to be made: i) the size and shape
of the accelerated body is not changed; ii) the change of the internal
energy of an accelerated classical gyroscope, caused by its increased
mass, can be generalized and it is valid for the internal energy of
any particles at rest in the accelerated system.

These assumptions added to the earlier ones form the foundation of
the present approach. The first consequence of the present approach
is that in the accelerated isolated subsystem, moving with a constant
speed $\vec V$ relative to the accelerator's system, the velocity
of the photon (or light) in the vacuum is $c'=c\sqrt{1-V^2/c^2}$,
where the velocity $c$ is the speed of the light in the accelerator's
system. In such a way the present approach seems to contradict both of
Einstein's axioms.

\subsubsection{The velocity of light emitted in a moving system}

Since the present approach are based upon the Newtonian model, the
velocities in different inertial frames are expressed according to
that model. The velocity of the photon $\vec c\,'$, emitted in the
accelerated system, will be $\vec c\,'+\vec V$ in the accelerator's
system. In case, when the photon is emitted in the direction of the
speed $\vec V$, its speed is $c'+V$ and in the opposite direction
it is $c'-V$. At the value $V=c/\sqrt{2}$ the speed of the photon
in the first case has its maximum value $c\sqrt{2}$ and in the
second case it has a zero velocity in the accelerator's system. If
$V>c/\sqrt{2}$ is true, then all photons, emitted in the accelerated
system, move away from the accelerator's system. Therefore if the
two systems are isolated, in case of $V>c/\sqrt{2}$ no emitted photon
from the accelerated system can reach the accelerator's system, while
the photons from the accelerator's system may reach the accelerated
system. However, in case of a semi-isolated subsystem, like an
elementary particle accelerated in a laboratory without leaving the
system of the Earth, the photon emitted by the accelerated particle
may behave differently. This may influence the description of the
Doppler effect.

\subsubsection{The velocity of light observed in a moving system}

Let us consider two systems: the original one fixed to an infinite
mass force-field and a finite mass moving subsystem, which was
accelerated in the original system to a constant velocity $\vec V$.
It is supposed that the two systems do not interact anymore, with
other words: they are isolated systems.

The present approach claims that the velocity of the photon, emitted
in the accelerated system and measured with the time unit defined
in the accelerator's system is $c'=c\sqrt{1-V^2/c^2}$. This can be
measured as $c'=L/\tau_L$, where $\tau_L$ is the time in the unit
defined in the accelerator's system, necessary for the light to
run the distance $L$ in the accelerated system. If the observer
of the accelerated system would like to measure the speed of the
same photon, using the time unit defined in its own system, the
$\widetilde c=L/\tau'_L$ relation has to be used. Because
$L/\tau'_L=(L/\tau_L)/\sqrt{1-V^2/c^2}$, the $\widetilde c=c$
is found.

If one accepts the present approach, the result of $\widetilde c=c$ is
erroneous, because the clock of the accelerated system is running more
slowly. If one would like to compare the different velocities, the
same units has to be used. If the measurer in the accelerated system
is aware that his clock is running more slowly than the clock of the
accelerator's system, he must to adjust his time unit to the unit of
the accelerator's system. Therefore he will get the $\widetilde c=c'$
result.

However, if the observer of the accelerated system believes what is
stated by Einstein's theory, namely that all inertial frames are
equivalent, then he must believe that its clock is running with the
same rate as that of the accelerator's system. Consequently he will
find that the measured velocity of the light $\widetilde c$ in the
accelerated system is the universal constant $c$. 

The problem is that both evaluations of the above measurement are
self-consistent.

It has to be emphasized that this is a crucial result, because
it shows how one's presumptions influence the evaluation of a
measurement.

\subsection{Summary}

There are two characteristic physical phenomena predicted by the
theory of special relativity: the increase of the mass of an
accelerated body and the time dilation. These phenomena were
proved by measurements of simple physical events, and therefore 
they may be considered as the clearest proof of the theory of
special relativity. This can be handled as the verification of
Einstein's axioms which lead to the present, universally accepted
idea that our world must be described in the four-dimensional
spacetime (the Minkowski space). In the present paper an attempt
is made to show that at least the above mentioned simple physical
phenomena can be explained in the classical, Newtonian space and
time.
 
To achieve this, first of all, the inertial mass is considered to
be the quantitative measure of the matter. As a such one it must be
independent on the coordinate frame it is observed from. Based upon
this assumption, a physical system has been found, which is believed
to be correctly described by the modified Newton's equation
(\ref{e1.2}). Supposing that the increased energy of an accelerated
particle is connected with the increase of its matter content (mass),
the relativistic dependence of the mass of the accelerated particle
has been reproduced. The reason of the changed mass of a body,
accelerated by an infinite mass force-field, is a real matter
transfer between the body and the accelerating force-field.

It has been found that certain inside physical processes of the
accelerated bodies become slower due to their increased mass and
as a consequence the time shown by a clock of an accelerated system
is less than the one shown by the same clock of the accelerator's
system. The lower rate of the accelerated clock is as real process,
independent of the system it is observed from, as the increase of
the mass of an accelerated particle. Because in the present approach
the frames of the accelerator's and accelerated systems are not
equivalent inertial frames, the extensively discussed twin paradox
of Einstein's theory does not exist (the increased mean lifetime
of an accelerated decaying particle is a real and asymmetric
phenomenon).

In order to reproduce the time dilation and the mass increase of
an accelerated particle, additional assumptions had to be made,
which of course may be disputable. Although these assumptions seem
to produce a coherent, self-consistent picture, unfortunately the
present approach is limited to the cases, when a finite mass
system is accelerated by a relatively infinite mass force-field,
while the Einstein's theory has a general validity.

\section{Combined motions}

Let us suppose that an infinite mass force-field of the system $S$
accelerates a finite mass subsystem $S'$ which has been at rest in
the system $S$ before the acceleration. In its final state the speed
of the subsystem $S'$ will be a constant velocity $\vec V$ and it will
become an isolated system. According to the present approach its mass
will be  $1/\sqrt{1-V^2/c^2}$ larger compared to its original rest
mass. In a next step, suppose that there is a secondary subsystem
$S''$ (for example a particle) at rest in the subsystem $S'$,
which was also at rest in the original system $S$ with a rest
mass $m_0$. This rest mass $m_0$ of the system $S''$ increases
to $m'_0=m_0/\sqrt{1-V^2/c^2}$ in the accelerated subsystem $S'$.
Let us suppose that the mass of the subsystem $S'$ and its
force-field acting upon the system $S''$ can be considered as
infinite ones compared to the mass $m'_0$. In this case the
description of the motion of the secondary subsystem $S''$ seems
to be possible.

The present approach claims that all of the masses of the system $S'$,
like the mass $m'$, are increased by a factor $1/\sqrt{1-V^2/c^2}$
and all of the internal energies are decreased according to the
relation (\ref{e3.2}). The decreased internal energy of a particle
with a mass $m'$ in the system $S'$ is supposed to be $m'c'\,^2$
(where $c'=c\sqrt{1-V^2/c^2}$ is the velocity of the light in the
subsystem $S'$), which is coherent with the expression (\ref{e3.2}).
This relation is supposed to be true for any parts, including the
force-fields, of the system $S'$. Therefore it seems to be logical
that the transferred energy due to the infinite mass force-field
of the subsystem $S'$ ought to be $c'\,^2{\rm d}m'$.  Based on
this, one may suppose that the form of the equation (\ref{e1.3})
for the particle accelerated by the infinite mass force-field of
the subsystem $S'$ ought to be
\begin{equation}
\vec v\,' {\rm d}(m'\vec v\,')=c'\,^2 {\rm d}m' \,\,\,  , \label{e4.1}
\end{equation}
where $\vec v\,'={\rm d}\vec r/{\rm d}t$ is the speed of the particle
in the subsystem $S'$ as it is seen from the original system $S$.

Multiplying the equation (\ref{e4.1}) by $m'$ and introducing the
$\vec p\,'=m'\vec v\,'$ one gets an equation ${\rm d}p'^2=
{\rm d}(m'^2c'^2)$. The same procedure with the equation (\ref{e1.3})
led to the ${\rm d}(m^2v^2)={\rm d}p^2={\rm d}(m^2c^2)$ and
because of $mc=m'c'$, the ${\rm d}p'^2={\rm d}p^2$ is true.
Taking into account that the subsequently accelerated system $S''$
in its initial state was at rest in both systems $S$ and $S'$, the
$p^2=p'^2 \rightarrow m^2v^2=m'^2v'^2$ leads to the relation
\begin{equation}
v'=v\sqrt{1-V^2/c^2} \,\,\, . \label{e4.2}
\end{equation}
The consequence of the relation (\ref{e4.2}) is that the equivalent
processes in the systems $S$ and $S'$ produce the same relations
for all of the equivalent velocities, than that of for the light
(\ref{e3.4}). However, if the observer of the subsystem $S'$ believes
that its clock and the equivalent clock of the system $S$ are
running with the same rate, then measuring the $v'$ by using the
clock of the system $S'$, the $v'=v$ result will be found.

Since in the initial state the particle (or the secondary subsystem
$S''$) was at rest in the subsystem $S'$ with a mass $m'_0$, the
solution of the equation (\ref{e4.1}) is
\begin{eqnarray}
m' & = & \frac{m'_0}{\sqrt{1-v'^2/c'^2}}=\frac{m_0}{\sqrt{1-V^2/c^2}
         \sqrt{1-v'^2/c'^2}} \nonumber \\
   & = &\frac{m_0}{\sqrt{1-(V^2+v'^2)/c^2}} \,\,\,\  . \label{e4.3}
\end{eqnarray}

To describe the motion of the particle in the system $S'$, the
Newton's equation of type (\ref{e1.1}) would be necessary. First
of all, let us allow that the force $\vec F\,'$ acting in the
system $S'$ may be different. Applying the relation (\ref{e1.3})
for the Newton's equation in the system $S'$, one will have an
equation $\vec F\,'{\rm d}\vec r\,'=c'^2 {\rm d}m'$. Combine it
with the relations (\ref{e1.6}) and (\ref{e3.4}) the result is
\begin{eqnarray}
\vec F\,'=\vec F \sqrt{1-V^2/c^2} \,\,\, , \label{e4.4}
\end{eqnarray}
because the distances are supposed to be the same in all
accelerated systems: ${\rm d}\vec r={\rm d}\vec r\,'$. This
result is not surprising, since the force supposed to be
related to the internal energy $E'_I=E_I\sqrt{1-V^2/c^2}$.
Therefore the final form of the Newton's equation in the
subsystem ought to be
\begin{eqnarray}
\frac{{\rm d}(m'\vec v \,')}{{\rm d}t}= \vec F\,'
   \,\,\, . \label{e4.5}
\end{eqnarray}

If one would like to get the equation (\ref{e4.5}) expressed
with time shown by the clock of the accelerated system $S'$,
the following form is found:
\begin{eqnarray}
\frac{{\rm d}(m'\vec v \,')}{{\rm d} t'}= \vec F
   \,\,\, , \label{e4.6}
\end{eqnarray}
because ${\rm d}t'={\rm d}t\sqrt{1-V^2/c^2}$. If an observer
in the system $S'$ believes that his frame is an inertial frame,
then using the Einstein's theory he believes that the mass
(for example the mass of a proton) and the time (a given frequency
of a given atom) units are the same than those of the system
$S$ ($m'=m$, $t'=t$). Based on the same belief, the measured
$\vec v\,'$ with the clock of the system $S'$ is also believed to
be equal to =$\vec v$ (see the explanation followed the expression
(\ref{e4.2})). Consequently for such an observer the equation
(\ref{e4.6}), valid in the system $S'$, is the same than the
corresponding Newton equation in the system $S$. Therefore this
observer of the system $S'$ believes that his system is an inertial
frame.

Although the frequency of a photon emitted by an atom is described
by the quantum mechanics, where instead of the Coulomb force the
Coulomb potential is needed, this potential in the system $S'$
supposed to be $\sqrt{1-V^2/c^2}$ less than that of the system $S$. 
According to the quantum mechanical solution, the energy of the
emitted photon between two energy levels is proportional to the
mass of the electron and to the square of the strength of Coulomb
potential. The increased mass of the electron and the decreased
Coulomb force in the system $S'$ will lead to the relation
(\ref{e3.1}). This indicates that the equations (\ref{e4.1}),
(\ref{e4.4}), and (\ref{e4.5}) may be correct.

Now simplifying the naming of the accelerated and the subsequently
accelerated systems $S'$, $S''$ and name all of them simply as systems.
Collaterally suppose that all of them became isolated systems. If one
had a particle at rest in the system $S$ with a mass $m$, this particle
is also at rest in both of the the system $S'$ accelerated to a constant
velocity $\vec V$ and in the system $S''$, subsequently accelerated by
the system $S'$ to a constant velocity $\vec V\,'$. The final mass $m''$
of the original particle in the system $S''$, according to the equation
(\ref{e4.3}), becomes
\begin{eqnarray}
m''=\frac{m'}{\sqrt{1-V'^2/c'^2}}=
  \frac{m}{\sqrt{1-(V^2+V'^2)/c^2}} \,\,\,\  . \label{e4.7}
\end{eqnarray}

One may introduce the concept of the mass-density of an isolated system,
which is proportional to its full mass or to the mass of any given parts
of it. The relation between the mass-densities of the systems $S_i$ and
$S_j$ can be characterized with a ratio $f(S_i,S_j)$ of the mass-densities
of the systems $S_i$ and $S_j$. For example $f(S,S')=\sqrt{1-V^2/c^2}$ and
$f(S,S'')=\sqrt{1-(V^2+V'\,^2)/c^2}$. The changed mass-density of the
systems $S''$ ought to lead to a further slowdown of the velocity $c''$
of photons and the time $\tau''$, shown by the clock of the system $S''$.
The internal energy $E''_I$ and the frequency of the emitted photons in
the system $S''$ are also less:
\begin{eqnarray}
c'' & = & c f(S,S'')\,\,\, ,\,\,\, \tau''=\tau f(S,S'') \,\,\, , \nonumber \\
E''_I & = & E_I f(S,S'')\,\,\, ,\,\,\, \nu''= \nu f(S,S'') \,\,\, .
    \label{e4.8}
\end{eqnarray}

The relation (\ref{e4.3}) clearly shows that inequalities $V \leq c$,
$V' \leq c'$, and $V^2+V'^2 \leq c^2$ must be satisfied. The particle
accelerated in the system $S'$ have a velocity $\vec v=\vec V+\vec V\,'$
in the original system $S$. If $\vec V'$ has the same direction as the
$\vec V$, then $v=V+V'$. In case of $V=c/\sqrt{2}$ for example, the
value $V'\leq c/\sqrt{2}$ is allowed and this may lead to a $v>c$
value. This means that according to the present approach even a material
body can be accelerated by subsequent accelerations to a speed larger
than the speed of light in the system $S$.

There is one special case for the secondary acceleration: the case
of $\vec V=-\vec V'$. The resulting secondary system will not move
relative to the original system. On the other hand, the equivalent
particles will have different masses and consequently the equivalent
clocks of the systems will run with a different rate. Since the
original and the secondary systems do not move relative to each
other, the different rate of their clocks, the different velocities
of their emitted light, and the supposed equality of the corresponding
distances easily can be checked.

An important consequence of the subsequent accelerations is that
according to the present approach the masses of the equivalent particles
and the rates of the equivalent clocks of isolated systems may be
independent on their relative velocity.

\section{Motion in a central force-field}

The simplest system for what solvable equations can be derived is the
motion of a pointlike body moved by a central force, supposed not to
be influenced by the pointlike body. The force acting on the pointlike
body must have a coordinate dependence $\vec r/r^3$, where $\vec r$
defines the position of the pointlike body in a coordinate frame,
fixed to the origin ($r=0$) of the central force.

\subsection{Gravitational interaction}

The form of the gravitational force according to our present knowledge
is $\vec F = \gamma M m \vec r/r^3$, where  $M$ and $m$ are the masses
of the two interacting pointlike bodies and $\gamma$ is a constant.
One supposes that $M>>m$ and therefore the possible change of $m$
practically does not change the force $\vec F$, which can be written
in the form $\vec F = -\alpha c^2 m \vec r/r^3$, where $\alpha =
\mid \gamma \mid M/c^2$. The validity of this setup for a planet as a
pointlike particle in the solar system, which moves in an infinite mass
force-field of the sun, is questionable, because the mass of a planet
may be non-negligible relative to the mass of the gravitational
force-field. However, the velocities of the planets are much less than
the velocity of the light and therefore, as a first approach, the
gravitational interaction may be approximated as a local and
instantaneous one.

In cases of central forces the introduction of the spherical coordinate
frame is the best choice. Inserting the last form of the force $\vec F$
into the relation (\ref{e1.3}) one gets the following equation:
\begin{eqnarray*}
{\rm d}m = - \alpha m\frac{1}{2}\frac{{\rm d}(r^2)}{(r^2)^{3/2}} =
             \alpha m{\rm d} \frac{1}{r} \,\,\, . 
\end{eqnarray*}
This equation, using the variable $\rho=1/r$, can be solved and the
result is
\begin{equation}
m = m_{in} \exp^{\alpha \rho -\alpha \rho_{in}} \,\,\, .  \label{e2.1}
\end{equation}
where $m_{in}$ and $\rho_{in}$ are the initial values of $m$ and $\rho$,
respectively, of the lighter body.

Returning to the expression (\ref{e1.5}), the actual speed $v$ of the
body of mass $m$ can be expressed using the mass $m$ as:
\begin{eqnarray}
v^2 = c^2 [ 1 - \frac{m_{in}^2}{m^2} (1-\beta_{in}^2)]
      \,\,\, , \label{e2.2}
\end{eqnarray}
where $\beta_{in} = v_{in}/c$ and $v_{in}$ is the speed of the lighter
body in the initial state.

Because the force is central, it has no angular component, and
consequently the $\varphi$ component of the Newton's equation
(\ref{e1.1}) becomes
\begin{eqnarray*}
\frac{1}{r}\frac{{\rm d}(mr^2{\rm d} \varphi/{\rm d} t)}{{\rm d} t}=
  \frac{1}{r}\frac{K}{{\rm d} t}=F_\varphi=0
\end{eqnarray*}
and the angular momentum is conserved:
\begin{eqnarray}
K=mr^2\frac{{\rm d} \varphi}{{\rm d} t}=
  m_{in}r_{in}^2\frac{{\rm d} \varphi_{in}}{{\rm d} t}=K_{in}
  \,\,\, .  \label{e2.3}
\end{eqnarray}

The angular momentum conservation enables one to use the derivation
by the angle variable $\varphi$ instead of the time derivation:
\begin{eqnarray}
\frac{{\rm d}r}{{\rm d}t}=\frac{{\rm d}r}{{\rm d} \varphi}
  \frac{{\rm d} \varphi}{{\rm d} t}=\frac{{\rm d}r}{{\rm d} \varphi}
  \frac{K}{mr^2} \,\,\, .  \label{e2.4}
\end{eqnarray}

Since $v^2$ can be expressed by the time derivate of $r$ and $\varphi$
($v^2=({\rm d}r/{\rm d}t)^2+(r{\rm d}\varphi/{\rm d}t)^2$), using the
variable $\rho=1/r$ and the relation (\ref{e2.4}), the final form of the
equation (\ref{e2.2}) is:
\begin{equation}
\rho^2+{\rho '}^2 = \frac{m_{in}^2 c^2}{K^2}\Big(\frac{m^2}{m_{in}^2}+
                    \beta_{in}^2 - 1\Big) \,\,\, , \label{e2.5}
\end{equation}
where $\rho '={\rm d}\rho/{\rm d}\varphi$.

In the next step one has to use the expression (\ref{e2.1}) for the mass
$m$ and the equality $K=K_{in}$, and the final equation becomes
\begin{equation}
\rho^2+{\rho '}^2 = \frac{m_{in}^2 c^2}{K_{in}^2}(\exp^{2(\alpha \rho
   -\alpha \rho_{in})} + \beta_{in}^2 - 1) \,\,\, . \label{e2.6}
\end{equation}

If the relations $\alpha \rho<<1$ and $\alpha \rho_{in} << 1$ are true,
one can substitute the exponential part for its Taylor expansion.
Keeping only the relevant terms, the final result is
\begin{equation}
\rho^2+{\rho '}^2 \approx \frac{m_{in}^2 c^2}{K_{in}^2}
   (\beta_{in}^2 - 2\alpha \rho_{in} + 2 \alpha \rho + 2 \alpha^2 \rho^2)
   \,\,\, . \label{e2.7}
\end{equation}

The solution of this approximate equation is
\begin{eqnarray*}
\rho \approx \frac{1+{\rm e}\cos {\omega \varphi}}{\rm p} \,\,\, ,
\end{eqnarray*}
where $\omega^2=1-2m_{in}^2 c^2\alpha^2/K_{in}^2$.
  
The importance of this result is, that the mass change of the body, moving
in a central force-field, leads to a peri-helium shift within the present
approach. In order to reproduce the measured peri-helium shift, on the
right hand side of the equation (\ref{e2.7}) a term $6\alpha^2\rho^2$
ought to be instead of the term $2\alpha^2\rho^2$.

The expression (\ref{e2.2}) can be used to calculate the trajectory of
a light particle (photon) going near to a large mass. It is supposed
that the particle is coming from infinity and therefore $\rho_{in}=0$.
If the $\alpha \rho<<1$ condition is true, using the relevant
terms of the Taylor expansion of the exponential function, the
equation (\ref{e2.6}) becomes
\begin{equation}
\rho^2+{\rho '}^2 \approx \frac{m_{in}^2 c^2}{K_{in}^2}
   (\beta_{in}^2 + 2 \alpha \rho) \,\,\, . \label{e2.8}
\end{equation}
If the particle is a photon, then $\beta_{in}=1$ and the equation
(\ref{e2.8}) describes the classical deflection of the light
by the gravitation. However, the measured deflection is twice
larger. In order to get this value, one ought to have on the
right hand side of the equation (\ref{e2.8}) a $4 \alpha \rho$
term instead of $2 \alpha \rho$.

\subsubsection{Possible extension of the gravitational interaction}

The original idea was that the interaction between two bodies is
due to their common force-field. If these bodies are far apart,
both of them is supposed to have some kind of force-field and
the interaction between them arises when these fields start to 
overlap. If one of the bodies has an infinite mass and the finite
mass body moves inside of the fixed field of the infinite mass
body, the interaction supposed to be determined by the common
fields of the bodies. The force acting between the two bodies
are generally measured in a static situation, when the two
bodies do not move relatively to each other. However, if the
finite mass body is moving in the fixed force-field of the
infinite mass body, it is imaginable that some extra force
arises, because the field of the finite mass body also moves
relative to the fixed infinite mass force-field.

Returning to the modified Newton's equation (\ref{e1.2}) one may
realize that adding to the original force $\vec F_0$ an extra term
$\Delta \vec F_0$, perpendicular to the velocity of the particle,
will not change the energy transferred to the particle, therefore
the results (\ref{e1.5}), (\ref{e2.1}), and  (\ref{e2.2}) remain
the same. On the other hand, the extra force may have a $\varphi$
component, which may change the angular momentum conservation
(\ref{e2.3}). The extra force $\Delta \vec F_0$ ought to be
connected with the velocity of the particle and consequently
with the change of the mass of the particle. Let us try an
arbitrarily chosen form:
\begin{eqnarray*}
\Delta \vec F_0 = -((\frac{{\rm d} m}{{\rm d}\vec r}
   \times \vec v) \times \vec v) \,\,\, ,
\end{eqnarray*}
which satisfies the above mentioned properties.

Using the relation (\ref{e1.3}), the full force can be written as
\begin{eqnarray}
\vec F=\vec F_0-\frac{1}{c^2}((\vec F_0 \times \vec v)\times \vec v)
   \,\,\, . \label{e2.9}
\end{eqnarray}
This form of the extra term has a ''magnetic'' force structure and
therefore it will be used by this name.

If $\vec F_0$ is a central force, one can express it in the form:
$\vec F_0=F_0\vec r/r$ and the spherical components of the force
$\vec F$ are:
\begin{eqnarray*}
F_r = F_0 (1+r^2 \dot {\varphi}^2/c^2)) \,\,\,  , \,\,\,\,\,\,\,
   F_\varphi = -\frac{F_0}{c^2} r \dot r \dot \varphi \,\,\,  ,
\end{eqnarray*}
where $\dot \varphi ={\rm d} \varphi/{\rm d} t$ and
$\dot r ={\rm d} r/{\rm d} t$. Since one has a generally nonzero
$F_\varphi$ component, the corresponding equation is
\begin{eqnarray*}
\frac{1}{r} \frac{{\rm d}(m r^2 \dot \varphi)}{{\rm d}t} = \frac{1}{r} 
  \frac{{\rm d}K}{{\rm d}t}=-\frac{F_0}{c^2}\dot r r \dot \varphi \,\,\, .
\end{eqnarray*}
If $F_0$ is the gravitational force ($F_0=-\alpha m c^2/r^2$), this
equation can be solved and the result is
\begin{equation}
K = K_{in} \exp^{\alpha \rho_{in} - \alpha \rho} \,\,\,  . \label{e2.10}
\end{equation}

Inserting the expressions (\ref{e2.1}) and (\ref{e2.10}) into the
equation (\ref{e2.5}) one gets the following equation: 
\begin{equation}
\rho^2+{\rho '}^2=\frac{m_{in}^2 c^2}{K_{in}^2}\exp^{2\alpha \rho-
                  2\alpha \rho_{in}} \Big(\exp^{2\alpha \rho-
                  2\alpha \rho_{in}}+\beta_{in}^2-1 \Big)
       \,\,\, .  \label{e2.11}
\end{equation}
Supposing the same $\alpha \rho<<1$ and $\alpha \rho_{in} << 1$
relations, one may substitute the Taylor series for the exponential
functions. Keeping the relevant terms, finally one gets the equation
\begin{eqnarray}
\rho^2+{\rho '}^2 \approx \frac{m_{in}^2 c^2}{K_{in}^2}
(\beta_{in}^2-2\alpha \rho_{in}+2\alpha \rho+6\alpha^2 \rho^2)
   \,\,\, .  \label{e2.12}
\end{eqnarray}

The equation of the trajectory of a photon going near to a large mass,
because of the second exponential in the equation (\ref{e2.11}), 
becomes
\begin{equation}
\rho^2+{\rho '}^2 \approx \frac{m_{in}^2 c^2}{K_{in}^2}
   (1 + 4 \alpha \rho) \,\,\, . \label{e2.13}
\end{equation}

In both cases (peri-helium shift and the deflection of the light),
adding the ''magnetic'' component to the classical gravitation,
the experimental values can be reproduced within the present approach.
It has to be emphasized, however, that originally these values were
predicted much more elegantly by the general theory of relativity
with the pure gravitational force.

One has to note also, that the presence of the ''magnetic'' force
will increase the gravitational attraction in case of a non-zero
angular momentum, since
\begin{eqnarray*}
F_r & = & F_0 (1+r^2 \dot {\varphi}^2/c^2)) = F_0 (1+\frac{K^2}{m^2 r^2 c^2}) \\
    & = & F_0 (1+\frac{K_{in}^2}{m_{in}^2 c^2} \rho ^2i
          \exp^{4\alpha (\rho_{in}-\rho)})
  \,\,\, .
\end{eqnarray*}

\subsection{Coulomb interaction}

Let us suppose that a finite mass particle with a mass $m$ and charge $Z_2e$
moves in the force-field of a relatively infinite mass body with a charge
$Z_1e$ ($Z_1$, $Z_2$ are positive or negative integers, and $e$ is the
unit of the electric charge). The Coulomb force acting between them is
$\vec F = Z_1 Z_2 e^2 \vec r/r^3$, which can be written analogously to the
gravitational force as $\vec F = -\alpha c^2 m_{in} \vec r/r^3$, where
$\alpha = -Z_1 Z_2 e^2 / m_{in} c^2$ (the mass $m_{in}$ is the initial
mass of the light particle in the system fixed to the infinite mass).
Although one gets an $\alpha$ depending on the initial state, the Coulomb
force remains independent on the masses of the interacting particles.

Inserting this $\vec F$ force into the equation (\ref{e1.3}) using the
variables $\rho=1/r$ and $\rho_{in}=1/r_{in}$ ($r_{in}$ is distance of
the light particle from the center of the force in its initial state),
for the mass $m$ the following solution is found:
\begin{equation}
m=m_{in} (1-\alpha \rho_{in}+\alpha \rho). \label{e2.15}
\end{equation}

Following the method used in case of the gravitational interaction,
one gets the equation (\ref{e2.5}). Using the relation $K=K_{in}=
m_{in} r_{in}^2 \dot \varphi_{in}$ and the expression (\ref{e2.15}),
the final form of the equation for the Coulomb interaction becomes
\begin{eqnarray}
& \rho^2 (1-\frac{m_{in}^2 c^2 \alpha^2}{K_{in}^2})+{\rho '}^2 = \nonumber \\
&  \frac{m_{in}^2 c^2}
{K_{in}^2}(\beta_{in}^2 - 2\alpha \rho_{in} + \alpha^2 \rho_{in}^2 +
    (1 - \alpha \rho_{in}) 2 \alpha \rho) \,\,\,\ . \label{e2.16}
\end{eqnarray}
This equation has an analytic solution
\begin{equation}
\rho = \frac{1+a\cos {\omega \varphi}}{b} \label{e2.17} \,\,\, ,
\end{equation}
with the parameter values:
\begin{eqnarray*}
 & \omega ^2 & = 1 - \frac{m_{in}^2 c^2 \alpha ^2}{K_{in}^2} \\ 
 & a ^2 & = 1 + \frac{\omega^2}{1-\omega^2}
          \frac{\beta_{in}^2-2 \alpha \rho_{in} + \alpha^2 \rho_{in}^2}
          {(1-\alpha \rho_{in})^2} \\
 & b & = \frac{\omega^2}{1-\omega^2} \frac{\alpha}{1-\alpha \rho_{in}}
    \,\,\,\ .
\end{eqnarray*}

Before one would like to apply the above solution for the hydrogen
atom, it has to be noted that the present approach is not applicable
for that case. Although the mass of the proton may be considered
as a relatively infinite one compared with the mass of the electron,
however, it is not accidental that all over the paper it is
emphasized that the present approach is valid only in case, when the
acting force-field of the infinite mass also has an infinite mass.
Since the the charges of the proton and electron are the same, their
independent force-fields (if they are far apart) are supposed to
have comparable quantity of the matter, and the condition that the
electron moves in a infinite mass force-field is not true.

\section{The behaviour of the photon}

According to the theory of relativity the photon (or light) has
special properties compared to a material body: i) its velocity
$c$ in the vacuum is an universal constant in all inertial frames;
ii) it has a zero rest mass.

In the present approach it has been supposed that the actual mass
of a body represents its quantity of matter. Therefore a photon
with a mass $m_f$ is also supposed to be a material body with an
$m_f$ quantity of matter. Since the mass of the photons in most
cases are negligible relative to the masses of bodies interacting
with them, there is a possibility that the motion of a pointlike
photon in some cases may be described by the modified Newton's
equations (\ref{e1.2}) and (\ref{e1.3}).

\subsection{Gravitational redshift}

Although the effect of the gravitational redshift at present believed
to be described properly only by the general theory of relativity, the
present approach predicts a redshift without any difficulty. The
redshift appears when a heavy body (a star) emits a photon, which
leaves the body and detected in another isolated system infinitely
far from the star. If the systems do not move relative to each other,
then the redshift is supposed to be the pure gravitational redshift.

However, in the present approach one has to satisfy an additional
condition: the mass densities must be the same for both systems.
Let us suppose that the star and the observer's system do not move
relative to each other and their mass densities are equal.Suppose
that the photon is emitted on the surface of the star with a mass
$M$ and radius $R$ in the radial direction and it is observed very
far from the star. The solution is $m'_f=m_f\exp^{-\alpha/R}$,
based on the expression (\ref{e2.1}). The $m_f$ is the mass of
the emitted photon, and $m'_f$ is the mass of the observed photon
far from the star. In case of $\alpha/R<<1$ one gets a relation
$m'_f=m_f(1-\alpha/R)$. The emitted frequency $\nu$ and the
observed frequency $\nu'$ are related as
\begin{eqnarray}
\nu'=\nu (1-\alpha/R) \,\,\, .  \label{e5.1a}
\end{eqnarray}

In principle, if the material system of the emitter and the system
of the observer do not move relative to the each other, but their
mass densities are different, then the relation (\ref{e5.1a}) ought
to be different. It is clear that according to the present approach,
the equivalent frequencies in these two systems are not equal and
therefore in case of no gravitation ($\alpha=0$), the expression
(\ref{e5.1a}) is not true. Fortunately the gravitational redshift
can be measured in a more controlled case: if nuclear $\gamma$ quanta
moves upward on the earth, a gravitational redshift of its frequency
can be observed within the same system.

\subsection{Refraction}

The simplest case is, when a photon enters into a homogeneous medium
from a vacuum, where its velocity is $c$. The experiences show that
if the speed of the photon is $v<c$ in the medium, then the direction
of the motion may be changed too. This change of the direction is
called refraction and it is characterized by the ratio $n=c/v$. The
generalization of the refraction for cases of two media with the
speed of photons $v_1$ and $v_2$ is straightforward.

\subsubsection{Refraction in a medium at rest}

The refraction is an expression of the angles between the normal
to the surface (perpendicular to the surface) in the entering
point from outside (the angle $i$) and from inside (the angle $r$):
\begin{equation}
\frac{\sin i}{\sin r} = \frac{c}{v} = n \,\,\, .  \label{e5.1}
\end{equation}

One may suppose that a homogeneous medium interacts with the photon
and this interaction causes the slowdown of the photon. Following
Newton's considerations one may suppose that the interaction acts in
the direction of the normal to the surface, since the new medium is
manifested in this direction. Therefore no interaction is supposed
perpendicular to this normal. This is the reason, confirmed by the
experiments, that the refraction takes places in the plane defined
by the directions of the normal and the momentum of the incoming photon.
Because of the missing interaction, the momentum of the photon parallel
to the surface must be unchanged:
\begin{eqnarray}
m_f c \sin i = m'_f v \sin r \,\,\,  , \label{e5.2}
\end{eqnarray}
where $m_f$ is the mass of the photon outside and $m'_f$ is its mass
inside. The new element of this formula relative to Newton's idea
is that because of the interaction between the medium and photon,
a mass transfer supposed to be possible between them.

The experiences indicate that the energy of the photon remain the
same. If one supposes that the energy of the photon is always has a
form $E_f=v_f^2m_f$ in every system (at least in the present approach
this is true for accelerated systems), then this energy conservation
is expressed as $E_f=c^2m_f=v^2m'_f=E'_f$ and one gets an expression
$m'_f=m_fc^2/v^2$. Substituting this expression into the formula 
(\ref{e5.2}) one gets the rule of refraction (\ref{e5.1}).

If the order of the above elaboration is reversed one may say that
substituting the rule of refraction (\ref{e5.1}) into the equation
(\ref{e5.2}), supposing the energy conservation of the photon, one
gets that the energy of the photon in the medium is $E'_f=v^2m'_f$.

An interesting remark: Newton insisted on the idea, that the masses
of the particles are constants. Therefore he wanted to use the rule
of the refraction to define the speed of the light in the medium,
which in case of the constant masses gives the clearly wrong $v=nc$
result.

\subsubsection{Refraction in a moving medium}

Following the basic idea that the mass of the photon may be changed
during the entrance into an other medium, the above used momentum
conservation of the photon has to be modified. As a first approach
the increased mass of the accelerated medium is neglected. The
transferred mass $\Delta m_f=m'_f-m_f$ has an additional momentum
$\Delta m_f \vec V$ because of the motion of the medium and this
momentum has to be added to the momentum $m'_f\vec v$ of the photon
in the standing medium. The resulting momentum of the photon in the
moving medium will be expressed as
\begin{eqnarray*}
m'_f \vec v + \Delta m_f \vec V = m'_f (\vec v +\frac{\Delta m_f}
  {m'_f} \vec V) = m'_f (\vec v + \Delta{\vec v}) \,\,\,   ,
\end{eqnarray*}
where
\begin{eqnarray*}
\Delta{\vec v}=\frac{m'_f-m_f}{m'_f} \vec V =(1-\frac{m_f}{m'_f})\vec V=
               (1-\frac{v^2}{c^2})\vec V \,\,\,  .
\end{eqnarray*}

If one takes into account that this is an approximation, finally one
gets the expression how the velocity of the photon in the moving medium
($\vec v+\Delta \vec v$) is changed:
\begin{equation}
\Delta{\vec v} \approx (1 - \frac{1}{n^2}) \vec V \,\,\,   . \label{e5.3}
\end{equation}

This agrees with the contemporary measurements.

\subsection{The Doppler effect}

If a wave is spreading in a medium between a source and a spectator,
which move in the medium, the emitted and observed frequencies are
different. This is the Doppler effect. The precondition of this
phenomenon is that the velocities of the emitter and spectator
in the medium must be less than the speed of the wave in the medium.
In the simple case when the wave is emitted by a source, moving with
a velocity $\vec V$ relative to the medium in which the observer is
at rest, the Doppler effect is expressed as
\begin{eqnarray*}
\widetilde \nu = \frac{\nu'}{1-\vec V \vec v/v^2} \,\,\, ,
\end{eqnarray*}
where $\widetilde \nu$ and $\nu'$ are the observed and emitted
frequencies, respectively, and $\vec v$ is the speed of the wave in
the medium.

\subsubsection{Doppler effect of the light waves}

If the wave is the light emitted by an atom, accelerated to a constant
speed $\vec V$ by an infinite mass force field of the observer's system,
the earlier formula becomes
\begin{eqnarray}
\widetilde \nu = \frac{\nu'}{1-\vec V \vec c/c^2} \,\,\,  . \label{e5.4}
\end{eqnarray}

The application of the formula (\ref{e5.4}) is based upon the supposed
property of the light that its speed in the observer's system is $c$.
However, according to the present approach, in the system of the source
the speed of the light is only $c'=c\sqrt{1-V^2/c^2}$. In case of an
accelerated atom one can get rid of this problem if the atom is
treated as moving source of a new wave, generated with a frequency
$\nu'$ and spreading with a speed $\vec c$ in the observer's system.
This interpretation means that the accelerated system moves inside the
matter of the accelerator's systems and it must be only a semi-isolated
system, which does not interact anymore with the original system. 

The expression (\ref{e5.4}) contradicts to the prediction of the
theory of relativity proven by measurements based upon the above
type of emission of a photon:
\begin{eqnarray}
\widetilde \nu = \nu \frac{\sqrt{1-V^2/c^2}}{1-\vec V \vec c/c^2} =
     \nu f_D  \,\,\, ,  \label{e5.5}
\end{eqnarray}
where $\nu$ is the emitted, $\widetilde \nu$ is the observed frequencies
and $f_D=\widetilde \nu/\nu$ is the Doppler factor.

The measurement detects the shift between the measured frequency
$\widetilde \nu$ and the supposed emitted frequency $\nu$. The chosen
frequency $\nu$ used to be some known frequency (for example a given
atomic or nuclear frequency). The theory of relativity claims
that i) the velocity of the light in vacuum is the universal constant
$c$ in all inertial frames; ii) the frequency, produced by a certain
process, is the same in every inertial frames. However, in the present
approach the emitted frequency $\nu'$ of a given process in the accelerated
system will be less than the corresponding frequency $\nu$ in the
accelerator's system (\ref{e3.1}). Since the Doppler factor $f_D$ is
the ratio of the measured frequency $\widetilde \nu$ and the frequency
$\nu$ of the equivalent physical process in the observer's system, the
$\nu'=\nu\sqrt{1-V^2/c^2}$ (relation (\ref{e3.1})) has to be substituted
into the formula (\ref{e5.4}) and the relativistic formula (\ref{e5.5})
for the Doppler factor $f_D$ is immediately reproduced.

\subsubsection{Doppler effect of photons}

If one handles the photons as particles, the $\widetilde m_f$ is the
mass of the photon, measured in the system $S$ and $m'_f$ is the mass
of the emitted photon in the system $S'$. Using the rules of quantum
theory $h\nu'=c'\,^2m'_f$ and $h\widetilde \nu= c^2\widetilde {m_f}$,
the expression (\ref{e5.4}) can be written as $\widetilde {m_f}=
m'_f(1-V^2/c^2)/(1-\vec V\vec c/c^2)$. Since the mass  $m'_f$ of the
photon emitted in the system $S'$ and the mass $m_f$ of the equivalent
photon in the system $S$ are related as $m'_f=m_f/\sqrt{1-V^2/c^2}$,
one gets the relation
\begin{eqnarray}
\widetilde m_f=m_f\frac{\sqrt{1-V^2/c^2}}{1-\vec V \vec c/c^2}
    \,\,\, .  \label{e5.6}
\end{eqnarray}

According to the present approach the speed of the photon, emitted in
the system $S'$ and treated as a particle, would have the velocity
$\vec c\,'+\vec V$, when it arrives into the system $S$. On the other
hand, one has to suppose that in the system $S$ it must move with a
velocity $c$. Therefore when the photon enters the matter (whatever
it means) of the system $S$, some kind of adaption to the system $S$
must take place. In principle the basic difference between the system
$S'$ and $S$ is their different mass-densities. One may suppose that
the adaptation is connected with some kind of adjustment of the
mass of the photon. The photon is a special kind of a particle,
because i) its mass may have any value (contrary to the elementary
particles); ii) according to the experiences, in any system its
velocity must be the a constant in every direction, depending on
the mass-density of the system. Since the isolated system $S'$
is supposed to move with a speed $\vec V$ in the system $S$, the
emitted photon will arrive into the system $S$ with a momentum
$m'_f(\vec c\,'+\vec V)$. This photon ought to interact with the
matter of the system $S$ and due to this interaction its mass
and momentum (including both the size and direction) has to be
changed in order to become a photon satisfying the system $S$.
The only freedom for this procedure is the change of the mass and
the direction of the velocity of the photon. Otherwise one cannot
explain the experienced effect: the photon within the system $S$
in every direction will have a velocity $c$ independent on the
motion of the source of the photon.

If the velocity $\vec c\,'$ of the emitted photon is perpendicular to
the velocity $\vec V$, then the relations $\vec c\,'+\vec V=\vec c$
and $\vec V \vec c=V^2$ are true. As a consequence, the relation
(\ref{e5.6}) will give a $\widetilde m_f=m'_f$ value, which shows
that if the photon, emitted in the system $S'$, would have a
velocity $\vec c=\vec c\,'+\vec V$ in the system $S$, no mass
adaptation process is necessary. In this case one may suppose that
no interaction arises between the photon and the matter of the
system $S$ and because of the lack of interaction, no mass-transfer
is taking place. Therefore it is understandable that the mass $m'_f$
of the photon is preserved while it enters the system $S$. The other
consequence of the lack of the interaction is, that in this special
case the direction of the speed of the photon must be conserved too.

If one supposes that the interaction between the entering photon and
the matter of the system $S$ is directed to the velocity $\vec V$ of
the system $S'$ relative to the system $S$, one has some guidance.
Let us introduce the components of $\vec c$ and $\vec c\,'$ parallel
and perpendicular to the speed $\vec V$: $c_\parallel$, $c'_\parallel$
and $c_\perp$, $c'_\perp$, respectively. The momentum conservation
$m'_f c'_\perp = \widetilde m_f c_\perp$ of the perpendicular
components is the consequence of the missing interaction in this
direction. This momentum conservation can be written by using the
relation $m'_f=m_f/\sqrt{1-V^2/c^2}$ as
\begin{eqnarray}
m_f c'_\perp = \widetilde m_f c_\perp \sqrt{1-V^2/c^2}
    \,\,\, . \label{e5.7}
\end{eqnarray}

In order to reproduce the relation (\ref{e5.6}) of masses based upon the
wave-like behaviour of photons, one must suppose that the relation for
the parallel components $c_\parallel$ and $c'_\parallel$ must be the
following one:
\begin{eqnarray}
m_f c'_\parallel = \widetilde m_f (c_\parallel-V) \,\,\, . \label{e5.8}
\end{eqnarray}
It has to be admitted that no solid explanation was found to justify the
validity of the relation (\ref{e5.8}).

Using the relations (\ref{e5.7}) and (\ref{e5.8}) one gets the
following expressions:
\begin{eqnarray*}
&&\widetilde m_f^2 [c_\perp^2 (1-V^2/c^2) + c_\parallel^2 + V^2-
   2 V c_\parallel] = \\ 
&&\widetilde m_f^2 [c_\perp^2 + c_\parallel^2 -(c^2-c_\parallel^2)
  V^2/c^2 +V^2 -2 V c_\parallel] = \\
&&\widetilde m_f^2 [c^2+V^2 c_\parallel^2/c^2-2 Vc_\parallel]= \\
&&\widetilde m_f^2 c^2 (1 - \vec V \vec c/c^2)^2 =
   m_f^2 c'^2 \,\,\, ,
\end{eqnarray*}
which reproduces the expression (\ref{e5.6}), because
$c'\,^2=c^2(1-V^2/c^2)$.

Although no argument was found why the relation (\ref{e5.8}) for
the parallel components of the momenta is true, using it one may
handle the case when the accelerated and the original systems ($S'$
and $S$, respectively) are absolutely isolated ones (they are far
apart and no matter exists between them). Contrary to the theory of
relativity, the present approach in this case includes the possibility
that the photon emitted in the accelerated system cannot reach the
original system. As it is mentioned before, if the velocity $\vec V$
of the accelerated system is larger than $c/\sqrt{2}$, no photon
emitted in that system $S'$ can reach the system $S$ and therefore
no Doppler effect can be observed. When the system $S'$ approaching
the system $S$, then in cases of $V>c/\sqrt{2}$ all of the photons,
emitted in the system $S'$ may reach the system $S$. Applying the
relation (\ref{e5.8}) to a photon with a velocity $c'_\parallel=-c'$,
one gets the equation $-m_fc'=\widetilde m_f(c_\parallel-V)$ which has
a solution only in case of $c_\parallel=-c$: $m_fc'=\widetilde m_f(c+V)$,
because the masses must be positive. This is an unexpected result, since
the supposed interaction between the matter of the system $S$ and
the incoming photon reverses the direction of the velocity of the
incoming photon. The result for the masses is
\begin{eqnarray*}
\widetilde m_f=m_f\sqrt{\frac{1-V/c}{1+V/c}} \,\,\, ,
\end{eqnarray*}
which is the same than that of the theory of relativity.

\subsubsection{The possible generalization of the Doppler effect}

The generalization of the wave model seems to be trivial. Consider
a system $S'$, which was subsequently accelerated starting from the
system $S$ (see the section of the ''Combined motion''). In this
case the mass-density of the system $S'$ and its velocity $\vec V$
in the system $S$ are not related. The frequency $\nu'$ of the photon,
emitted in the system $S'$ is related to the equivalent frequency
$\nu$ of the system $S$ as $\nu'=\nu f(S,S')$ (see the expression
(\ref{e4.8})) and the formula (\ref{e5.5}) is changed to the
\begin{eqnarray}
\widetilde \nu = \nu \frac{f(S,S')}{1-\vec V \vec c/c^2} =
     \nu f_D  \,\,\, ,  \label{e5.9}
\end{eqnarray}
where $\vec c$ is the speed of the light, measured in the system $S$.
Of course, if the system $S'$ accelerated directly to a speed $\vec V$
by the infinite mass force-field of the system $S$, the mass-factor is
$f(S,S')=\sqrt{1-V^2/c^2}$ and the expressions ({\ref{e5.5}) and
(\ref{e5.9}) for the Doppler effect are the same.

The resulting mass $\widetilde m_f$ of the photon coming from system
$S'$, and the equivalent mass $m_f$ in the system $S$ are related as
\begin{eqnarray}
\widetilde m_f=m_f\frac{f(S,S')}{1-\vec V \vec c/c^2}
    \,\,\, ,  \label{e5.10}
\end{eqnarray}

The same result comes from the relations (\ref{e5.7}) and (\ref{e5.8}),
if the velocity $\vec V$ is interpreted as the relative motion of the
system $S'$, seen from the coordinate frame fixed to the system $S$,
because $c'=cf(S,S')$ (see the relations (\ref{e4.8})).

Using the relations (\ref{e5.7}), (\ref{e5.8}), and the generalized
relation (\ref{e5.10}) one can define the direction of the photon
within the system $S$:
\begin{eqnarray*}
c'_\perp =  c_\perp \frac{f(S,S')\sqrt{1-V^2/c^2}}{1-\vec V \vec c/c^2}
   \,\,\, , \,\,\, c'_\parallel = \frac{f(S,S')(c_\parallel-V)}
   {1-\vec V \vec c/c^2} \,\,\, .
\end{eqnarray*}
It is extremely interesting to note that the ratio 
\begin{eqnarray}
\frac{c'_\parallel}{c'_\perp}=\frac{c_\parallel-V}{c_\perp\sqrt{1-V^2/c^2}}
    \label{e5.11}
\end{eqnarray}
depends only on the relative speed of the systems $S$ and $S'$ (the $V$
is measured in the system $S$).

The theory of relativity claims that if in a system $S$ an $S'$ subsystem
moves with a constant velocity $\vec V$ and a particle moves with a
velocity $\vec u$, then an observer in the system $S'$ will find that
the particle in the system $S'$ moves with a velocity $\vec u\,'$, whose
$u'_\parallel$ and $u'_\perp$ components parallel and perpendicular to
$\vec V$, respectively, are expressed as:
\begin{eqnarray*}
u'_\parallel=\frac{u_\parallel-V}{1-\vec V \vec u/c^2} \,\,\, ,  
 \,\,\,\,  u'_\perp=u_\perp \frac{\sqrt{1-V^2/c^2}}{1-\vec V \vec u/c^2}
 \,\,\,  ,
\end{eqnarray*}
where $u_\parallel$ and $u_\perp$ are the corresponding components
of $\vec u$. Using the $\vec c$ and $\vec c\,'$ instead of $\vec u$
and $\vec u\,'$ the ratio, defined by Einstein's theory is the same as
that of the relation (\ref{e5.11}).

\subsubsection{Possible consequences}

A consequence of the generalized Doppler effect (if it is valid at
all) expressed by formula (\ref{e5.9}) or (\ref{e5.10}) is that
in case of the zero relative speed between the systems $S$ and $S'$,
the energies $\widetilde m c^2$ and $m'c'\,^2$ of the photons are
equal. This indicates that in cases when the two systems do not move
relative to each other, the energy of the photon, entering from
one system into the other one, is conserved. A similar conservation
of the energy of a photon was supposed in case of the refraction
($m'v^2=mc^2$).

A second consequence, however, is much more important. If the
generalized formula (\ref{e5.9}) is true, the astronomical
measurements based upon the Doppler effect becomes questionable.

Both of the Newtonian picture and the theory of relativity claim
that the Doppler shift determines in a unique way the relative
motions of two isolated systems (of course in the universe the
gravitational effects, mainly the gravitational red-shift, in
any model have to be taken into account). Contrary to that, in
the present approach for systems, moving with a constant velocity
relative to each other, the Doppler shift may be used to
determine the speed of the relative motion only in cases,
when one of the systems was accelerated by the infinite mass
force-field of the other system. If the present approach may be
generalized for the subsequently accelerated systems, when 
the mass-densities and the constant relative speed of the systems
are not related, the Doppler factor depends on both of them
(see the formula (\ref{e5.9})).

In case of a rather artificial two-step subsequent acceleration,
when $\vec V\,'=-\vec V$ (it is possible only for cases
$V<c/\sqrt{2}$), the two systems will not move relative to
each other. The expression (\ref{e5.9}) gives a Doppler factor
$f_D=f(S,S')=\sqrt{1-2V^2/c^2}$ (see the expressions (\ref{e4.8}))
for the light emitted in the secondary accelerated system and
measured in the original system. Therefore, contrary to any earlier
theories, the present approach allows the creation of two systems with
zero relative speed and different frequencies of the equivalent
photons. It is clearly the effect of the different masses of the
equivalent particles of the two systems. The problem starts when
these systems are far apart from each other. Observing in one of
the systems the light, emitted from the other system, a Doppler
effect would be found. Using the Newton's or Einstein's theory,
this Doppler effect would be treated as a proof that the two systems
moves relative to each other. Depending on the used theory (Newton's
or Einstein's) even the speed of the relative motion can be defined,
while in the presented special case one knows that the systems do not
move relative to each other!

\subsubsection{Summary and conclusions}

The presented approach for the Doppler effect of the photons as
particles is mostly a phenomenological one, which includes some
reasonable physical consideration and some arbitrary momentum
relations accepted in order to reproduce the relativistic result
in case of a photon emitted by a finite mass system accelerated
by a relatively infinite mass force-field.

Let us suppose that there are two isolated material systems $S$
and $S'$, and $S'$ moves relative to the system $S$ with a velocity
$\vec V$ seen from the system $S$. Suppose that in the system $S'$
a photon is emitted with a speed $\vec c\,'$, where the velocity
$c'$ is measured by the clock of the system $S$. Since the present
approach is based upon the Newtonian physics, the speed of this photon
in the system $S$ ought to be $\vec c\,'+\vec V$ and one has to
suppose that after leaving the matter of the system $S'$, this
velocity is conserved. If this photon enters the matter of the
system $S$, according to the experiences, its velocity becomes $c$,
which is the velocity of the photon emitted in the system $S$. The
only way to explain this phenomenon is to suppose the existence of
an interaction between the incoming photon and the matter of the
system $S$. This form of the matter is probably a kind of a medium,
formed by the force-fields of the bodies the system $S$ composed of.
This medium must be also a material one and therefore in the
interaction process a mass exchange between the incoming photon
and the medium is possible. If the system $S$ is an extended body,
the mass of its medium relative to the mass of the photon may
be considered to be infinite.

The view that the photon moves in a medium is not unusual.
In case of a refraction one also have to suppose some
interaction between the incoming photon and the medium
in which the photon has a different velocity.

Of course the proper physical description of the Doppler
effect would require a stricter formulation of the supposed
interaction between the photon and medium. Without this, the
above phenomenology is just an indication that there is
an other possible explanation of the Doppler effect measured
in the controlled laboratory systems. The astronomical objects
of the Universe are probably do not belong to the isolated
systems accelerated by another infinite mass system, therefore
the present approach cannot be applied for their observed Doppler
effects. However, the present approach indicates that there may
be isolated systems for which the Doppler effect is different
from that of the theory of relativity and therefore the Doppler
shift does not allow to determine their relative speed with an
absolute certainty.

\section{Summary}

The following basic hypotheses are used:

i) our world is exclusively material and the quantity of the matter
(whatever it is) must be conserved;

ii) isolated material systems exist (there is no material connection
between them), which occupy final volumes of the space. This
assumption is an approximate one: in the Universe the gravitation
is supposed to act between the celestial objects.

Based upon these basic assumptions, it seems to be evident that
material systems generally ought to be described as a change of
their matter distribution in the space they occupy, including the
possibility that the size and shape of the occupied space also may
be changed. The description of a matter distribution most likely
ought to be based on some kind of field theories. However, if one
follows the instinctive perception that a material system is
composed of distinct interacting bodies, then the forces between
these bodies, generated by their common material force-field, in
principle ought to be handled as non-local and time-dependent ones.

Fortunately there are material systems, which can be approximated
as distinct bodies, whose motions are ruled by local and instantaneous
forces between them. These systems can be described successfully by
the Newtonian physics, which supposes that the masses (tacitly
believed to be the measure of the  matter content) of the distinct
bodies are constant. 

The simplest isolated system, which seems to satisfy the above
condition, is a pointlike body (particle) with a finite matter
content moved by a force, generated by an other body with a relatively
infinite matter content. If one takes seriously the hypothesis of
the exclusively material world, then the force ought to be also
material. In order to have a local and instantaneous force, one has
to suppose that the matter content of the force-field (the source
of the interaction) is also infinite relative to the finite matter
content of the particle.

It has been shown that for such system a modified Newton's equation
(\ref{e1.2}), defined in the Newtonian frame of space and time, can
be applied including the possibility of a matter transfer between
the particle and the force-field. This can be achieved by using the
following assumptions:

iii) The physical entity, characterizing the scalar measure of the
quantity of the matter, is the inertial mass of an isolated system,
perceived as a body. The inertial mass of such a body ought to
include the matter content of the force-fields connected with
it (inside and outside). Because of this, the inertial mass is
handled as a general measure of the matter content.

iv) Due to the action of the force, generated by a relatively infinite
mass force-field, the full energy $E$ of the particle increases by a
${\rm d}E$ value, proportional to the transferred mass ${\rm d}m$ as
${\rm d}E=c^2{\rm d}m$.

As a result the modified Newton's equation (\ref{e1.2}) produces a
mass increase of the accelerated particle equal to the one predicted
by Einstein's theory of special relativity. This result is achieved
on the basis of the Newtonian physics, although the Newtonian concept
that the masses of bodies moved by outside forces remain constant,
had to be abandoned.

It has been emphasized that there is a basic difference between the
concept of mass of the present approach and that of Einstein's: the
change of the mass, as a quantitative measure of the matter content
of a body, is a physical reality and it is the same in all coordinate
frames. If this is true, then the equivalence of all inertial frames
has to be ruled out.

Besides the increased mass of the accelerated particle, another
proved consequence of the Einstein's theory is the time dilation:
an accelerated clock is running more slowly compared to the same
clock of the accelerator's system. The present approach uses the
generalization of the changed properties of an accelerated
gyroscope due to its increased mass. The following new assumptions
had to be made:

v) the size and shape of a pointlike finite mass body is not
changed during its acceleration by an infinite mass force-field;

vi) the internal energy of the accelerated body, due to its
increased mass, becomes smaller (\ref{e3.2}).

Using the assumption v), it has been shown that due to the increased
mass, an accelerated mechanical clock is running more slowly
(\ref{e3.1}), than the equivalent clock of the accelerator's system
and the rate of the accelerated clock is equal to the one predicted
by Einstein's theory. Using assumption vi), an explanation has been
found that the same time dilation is valid for clocks, based upon
the frequencies of certain photons, emitted in the accelerated system.
In the present approach the slower rate of the accelerated clock is
as real process as that of the increased mass: it is independent on
the system observed from.

An important consequence of the changed internal energy of the
accelerated particles is that the velocity of the light, emitted
in the accelerated and observed from the accelerator's system, is
$\sqrt{1-V^2/c^2}$ smaller than the speed of the light in the
accelerator's system (\ref{e3.4}).

According to the present approach, the subsequent acceleration of
a finite mass system by a relatively infinite mass force-field of
a similarly accelerated system (see the section ''Combined
motions''), may lead to systems, whose relative mass-density
(the ratio of the masses of the same type of particles of the two
systems) is independent on the velocity of their relative motion.
Besides that, the subsequent accelerations may produce particles,
whose velocity in the in the original system is larger than the
speed of the light.

Up to this point, the present approach, based upon certain
assumptions, seems to be coherent and mathematically correct.

The present approach, although it is valid only for systems accelerated
by a relatively infinite mass force-field, excludes the Einstein's
principle that all inertial frames are equivalent and the velocity
of the light in the vacuum of every inertial frame is a universal
constant $c$. On the other hand, the present approach states that an
observer in such an accelerated system, using a clock of his own system
and believing that it is running with the same rate as the equivalent
clock of the accelerator's system, will find the speed of the light,
emitted and measured in his system, to be the same than that of the
one, emitted and measured in the accelerator's system. What is more,
if the observer of the accelerated system believes that the equivalent
masses of his and those of the accelerator's system are also equal
as it is stated by the Einstein's theory, he will find that all
equivalent processes in the accelerated system are the same as those
of the accelerator's system. All of these seem to justify the
Einstein's axioms. However, according to the present approach it is
a consequence of the erroneous presumptions that the equivalent
clocks at rest are running with the same rate and the equivalent
masses are equal in any inertial frames. This example serves as
a serious warning: the interpretation of a given measurement,
evaluated by exploiting some theoretical principles, may lead to
conclusions, which are erroneous in another approach.

It is interesting to note that allowing the change of the mass,
the Newton's equation leads to some peri-helium shift. If one
allows a ''magnetic force'', connected with the mass-transfer,
even the measured peri-helium shift and light deflection is
reproduced.

In a next step an attempt has been made to describe some
characteristic phenomena of the photon (the light) in different
systems. It was supposed that if the photon interacts with the
matter of a system, it satisfies the condition that the mass of
the photon is negligible relative to the mass of the matter of
the systems. The description of some phenomena requires reasonable
new assumptions. However, the description of the Doppler effect
is somehow more problematic, because even the simplest case
requires partly phenomenological assumptions in order to reproduce
the results of the Einstein's theory, which were proved by
measurements. The generalization of this phenomenological
description of the Doppler effect for the case of a single
accelerated system has led to a different result for systems,
whose mass-density is independent on their relative velocity.
This result, if it is correct, may question the reliability of
the astronomical measurements, based upon the Doppler effect of
the light coming from the objects far away in the Universe.

Here one has to return to a basic assumption that in the Newtonian
space and time isolated systems may exist in a vacuum (space with
no matter content), which may move relative to each other. If a
photon as a particle arrives from an isolated system to another
one, the present approach has to suppose that the photon somehow
interacts with the matter (or the medium) of the system it enters.
Due to this interaction an adaptation process takes place. As a
result the velocity of the photon within the system it enters
becomes the velocity defined by the mass density of the given
system. However, according to the present approach, the change
of the mass density of a system, accelerated by an infinite mass
force-field, is changing continuously. Therefore even in a system,
which is still interacts with the accelerator's force-field, the
velocity of the entering photon ought to be changed in accordance
with the actual mass density of the accelerated system. This can
be the explanation of the Michelson-Morley experiment: although
the Earth is not an isolated body in the Solar System and its mass
density depends on his actual velocity relative to the Sun (which
is approximated as an infinite mass system), the velocity of the
photon emitted by the Sun will be adapted to the actual mass density
of the Earth, when it enters into it. It is an other question that
there is no sharp division between the medium of the Sun and Earth,
therefore such type of adaptation ought to be a continuous process.

It has to be admitted that the above elaborated idea of a medium of
an isolated system, which is characterized in the present approach by
the mass density of the system, is a kind of a local ether (what
may be an internal force-field of the system), which may interact
with the photon.

\section{Subjective conclusions}

The starting point of the present approach is a firm conviction
that the existing world is exclusively material. One consequence
of this conviction is that if isolated (absolute or approximate)
material systems exist, in principle they ought to be described
as matter distributions in the space they occupy and their matter
content must be conserved.

Adding the idea that the Newtonian inertial mass is supposed to
be the physical entity expressing the quantitative measure of the
matter content, the probably most unambiguous experimental proof
of Einstein's theory, the increased mass of an accelerated particle,
is reproduced within the Newtonian frame of space and time.
However, there is an essential difference of the present and the
Einstein's interpretation of this phenomenon: the present approach
claims that this increased mass of the accelerated particle is a
real increase of its matter content and it is independent on the
coordinate frame observed from.

The Newtonian principle of the equivalence of all inertial frames is
just the consequence of the constancy of the masses of the distinct
bodies, therefore the possibility that these masses may be changed
due to an interaction, invalidates this conception.

The theory of special relativity is a general kinematical
description of the system of moving distinct bodies. It is
based on two axioms, one of them is the equivalence of all
inertial frames.

It has to be noted that the inertial frames in Newton's physics are
connected with the Newtonian space and the Galilean transformation.
In Einstein's physics the conception of the inertial frames is
connected with the Minkowski space and the Lorentz transformation.
In case of a single acceleration, the Lorentz transformation seems
to be valid in some way, however, in case of subsequent accelerations,
the relation between the equivalent masses and the rate of equivalent
clocks are independent on the relative velocity of the isolated
systems. Therefore the present approach contradicts both the Newton's
and Einstein's theories.

The second basically new feature is, that the present approach
makes questionable already the description of the world as a
system of distinct bodies. It is an other question, that the
examined special system is approximated as the motion of a
distinct body: otherwise no analytic solution could be found.

In Einstein's theory the two basic axioms led to the necessity of
the introduction of the four-dimensional spacetime. Unfortunately,
there is no room in this theory to find a conserved scalar physical
entity, which can be identified as the quantitative measure of the
matter content (amount of the matter). Of course one may state that
there is no need for such physical entity, it would be only an
unnecessary constraint. Although it has to be admitted that this may
be an acceptable opinion, the present author does not sympathize with
it.

The present approach is valid only in a special case, which belong
to the class for what the Newton's and Einstein's theories are valid.
The well-controlled special case of an accelerated system by a
relatively infinite matter force-field exists due to the human
activity. The possibility to create a secondary accelerated subsystem
is questionable. However, theoretically it may exist and the possible
consequences ought to be taken into account. Although these systems
are very much simplified, the same is true for the basic formulation
of Newton's and Einstein's theories.

The present approach is formulated in the Newtonian space and
time and the physical entity characterizing the matter content
is the changeable Newtonian mass. To the best knowledge of the
present author, this idea was not tested before. In the present
paper this idea was applied to describe some physical phenomena,
until now described properly only within the Einstein's theories,
and some experienced facts are reproduced. Of course there exist
thousands of other physical problems, which are described
successfully by Einstein's theory, however, because of the
embryonic state of the present approach, it cannot be expected
to achieve the same completeness.

Returning to the vision that in a general case our world ought
to be described as a matter distribution, one may ask about the
coordinate frame. If the Newtonian mass is really characterizes
the quantity of the matter, it may change the present views about
the existing world around us. Since this approach supposes,
contrary to the Einstein's model, a Newtonian space and time,
one may ask the question: are we really sure that we live in
the four-dimensional spacetime?

\begin{center}

\end{center}

\end{document}